\documentclass[%
 superscriptaddress,
 reprint,
 showpacs,preprintnumbers,
 nofootinbib,
 amsmath,amssymb,
 aps,
 prd,
 floatfix,
]{revtex4-2}

\usepackage{graphicx}
\usepackage{nameref}
\usepackage{xcolor}
\usepackage{lineno}
\usepackage{adjustbox} 

\usepackage{amssymb,amsmath,amstext,amsthm,amsfonts,amsfonts}
\usepackage[utf8]{inputenc}

\usepackage[caption=false]{subfig}

\usepackage[linesnumbered,ruled]{algorithm2e}
\usepackage{url}
\usepackage{soul}
\usepackage{bm}
\usepackage[normalem]{ulem}
\usepackage{tabularx}
\usepackage{multirow, makecell}
\usepackage{floatrow}
\usepackage{comment}

\usepackage{color, colortbl}
\definecolor{Gray}{gray}{0.9}
\definecolor{LightCyan}{rgb}{0.88,1,1}
\usepackage[first=0,last=9]{lcg}

\usepackage{amsmath}

\newcommand{\tbd}[1]{\textcolor{red}{#1}} 

\newcommand{\polystyrene}{PST} 

\newcommand{\supercube}{SuperCube}

\DeclareMathAlphabet{\mathsfit}{\encodingdefault}{\sfdefault}{m}{sl}
\SetMathAlphabet{\mathsfit}{bold}{\encodingdefault}{\sfdefault}{bx}{n}

\begin{document}

	\title{Additive manufacturing of a 3D-segmented plastic scintillator detector for tracking and calorimetry of elementary particles}

    \author{Tim Weber}
    \affiliation{ETH Zurich, Institute for Particle Physics and Astrophysics, CH-8093 Zurich, Switzerland}

    \author{Andrey Boyarintsev}
    \affiliation{Institute for Scintillation Materials NAS of Ukraine (ISMA), National Academy of Science of Ukraine (NAS), Nauki ave. 60, Kharkiv 61072, Ukraine}

    \author{Umut Kose}
    \author{Boato Li}
    \author{Davide Sgalaberna}
    \email[E-mail:] {davide.sgalaberna@cern.ch}
    \affiliation{ETH Zurich, Institute for Particle Physics and Astrophysics, CH-8093 Zurich, Switzerland}

    \author{Tetiana Sibilieva} 
    \affiliation{Institute for Scintillation Materials NAS of Ukraine (ISMA), National Academy of Science of Ukraine (NAS), Nauki ave. 60, Kharkiv 61072, Ukraine}

    \author{Siddartha Berns}
    \author{Eric Boillat}
    \affiliation{Haute Ecole Sp\'ecialis\'ee de Suisse Occidentale (HES-SO), CH-2800 Del\'emont, Route de Moutier 14, Switzerland}
    \affiliation{Haute Ecole d'Ing\'enierie du canton de Vaud (HEIG-VD), CH-1401 Yverdon-les-Bains, Route de Cheseaux 1, Switzerland}
    \affiliation{COMATEC-AddiPole, CH-1450 Sainte-Croix, Technopole de Sainte-Croix, Rue du Progr\`es 31, Switzerland}

    \author{Albert De Roeck} 
    \affiliation{Experimental Physics department, European Organization for Nuclear Research (CERN), Esplanade des Particules 1, 1211 Geneva 23, Switzerland}

    \author{Till Dieminger}
    \affiliation{ETH Zurich, Institute for Particle Physics and Astrophysics, CH-8093 Zurich, Switzerland}

    \author{Stephen Dolan}
    \affiliation{Experimental Physics department, European Organization for Nuclear Research (CERN), Esplanade des Particules 1, 1211 Geneva 23, Switzerland}

    \author{Matthew Franks}
    \affiliation{ETH Zurich, Institute for Particle Physics and Astrophysics, CH-8093 Zurich, Switzerland}

    \author{Boris Grynyov}
    \affiliation{Institute for Scintillation Materials NAS of Ukraine (ISMA), National Academy of Science of Ukraine (NAS), Nauki ave. 60, Kharkiv 61072, Ukraine}
    
    \author{Sylvain Hugon}
    \affiliation{Haute Ecole Sp\'ecialis\'ee de Suisse Occidentale (HES-SO), CH-2800 Del\'emont, Route de Moutier 14, Switzerland}
    \affiliation{Haute Ecole d'Ing\'enierie du canton de Vaud (HEIG-VD), CH-1401 Yverdon-les-Bains, Route de Cheseaux 1, Switzerland}
    \affiliation{COMATEC-AddiPole, CH-1450 Sainte-Croix, Technopole de Sainte-Croix, Rue du Progr\`es 31, Switzerland}

    \author{Carsten Jaeschke}
    \author{Andr\'e Rubbia}
    \affiliation{ETH Zurich, Institute for Particle Physics and Astrophysics, CH-8093 Zurich, Switzerland}
    





\begin{abstract}
\noindent

Plastic-scintillator detectors are devices used for the detection of elementary particles. They provide good particle identification with excellent time resolution, whilst being inexpensive due to the affordability of plastic materials. Particle tracking is achieved by segmenting the scintillator into smaller optically-isolated 3D granular sub-structures which require the integration of multiple types of plastic materials as well as several thousands of tiny holes through a compact volume of several cubic meters.
Future particle detectors necessitate larger volumes, possibly with even finer segmentation. However, manufacturing such geometries with current production strategies is challenging, as they involve time-consuming and costly fabrication processes, followed by the assembly of millions of individual parts. The difficulty in scaling up such a workflow can be addressed by additive manufacturing, enabling the construction of complex, monolithic geometries in a single operation. 
This article presents the fabrication of the first additive manufactured plastic scintillator detector, capable of 3D tracking elementary particles and measuring their stopping power. Its performance is comparable to the state of the art of plastic scintillator detectors. This work paves the way towards a new feasible, time and cost-effective process for the production of future plastic-based scintillator detectors, regardless their size and difficulty in geometry.

\end{abstract}

\maketitle

\clearpage
\newpage

\section{Introduction}
\label{sec:introduction}

Plastic scintillator (PS) detectors,
invented in the early 1950s \cite{PhysRev.80.474},
are widely used in the detection of elementary particles in high-energy physics (HEP) \cite{Amaudruz:2012esa,Aliaga:2013uqz,MINOS:2008hdf,Joram:2015ymp,TheATLASCollaboration_2008}, 
nuclear physics \cite{PERDIKAKIS2012117}, 
astroparticle physics \cite{thepierreaugercollaboration2016pierre,CHANG20176}, 
as well as in many applications like muon tomography \cite{muon-tomography-pyramid}, 
proton computed tomography for hadron therapy \cite{pCT-1}, 
fast-neutron detection \cite{RCHaight_2012,LANGFORD201578} 
and non-destructive imaging \cite{SEKI2017148}.

\begin{figure*}[htb]
    \includegraphics[width=.9\linewidth]{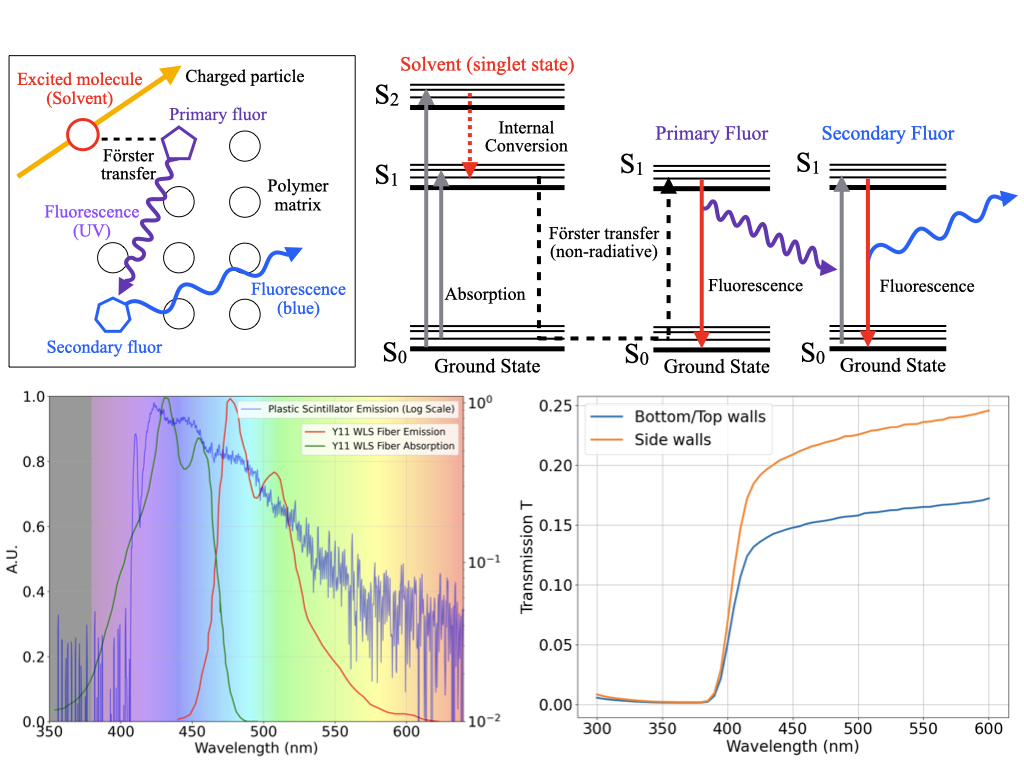} 
    \caption{ 
    Top: representation of the scintillation process in PS doped with a primary and a secondary fluor when the material is crossed by a charged particle.    
    Bottom: 
    emission spectrum of a 3D printed
    PS sample (blue) taken with a spectrophotometer, Kuraray Y11 WLS fiber absorption (green) and emission (red) spectra. 
    \label{fig:scintillation-process}  
    \label{fig:3dprint-cube}  
    }
\end{figure*}

By measuring the energy loss of a particle and tracking its path in the detector, it is possible to identify the type of interacting particle, reconstruct its momentum based on range, measure its original energy using calorimetry, and determine its electric charge if the setup is immersed in a magnetic field.
Another important feature of PS detectors is their unique ability to provide an extremely fast response, with time resolution in the sub-nanosecond range.
PS detectors are typically made of 
long scintillating bars with O(cm) granularity for time-of-flight detectors 
\cite{Betancourt2017,Korzenev:2021mny}, neutrino active targets with tonne-to-kilotonne scale mass \cite{Amaudruz:2012esa,Aliaga:2013uqz,MINOS:2008hdf},
sampling calorimeters made by layers of segmented PS alternated with heavier materials like iron and lead \cite{Allan:2013ofa}, 
or scintillating optical fibers \cite{Joram:2015ymp} with a diameter down to $250 ~\mu \text{m}$. 
\newline A PS is an organic material composed of a mixture of carbon and hydrogen-based molecules, typically polystyrene (\polystyrene) or polyvinyltoluene (PVT). Molecules of an activator like p-terphenyl (pTP), 2,2-p-phenylene-bis(5-phenyloxazole) (POPOP), 2,5-diphenyloxazole (PPO) are introduced into the polymer at a level between a few per mille and percent by weight. 
The scintillation mechanism consists of a few steps, highlighted in Fig.~\ref{fig:scintillation-process}. 
When a charged particle propagates through a PS, the molecules of the polymer matrix get excited. A fast short-range resonant non-radiative dipole-dipole interaction, called Förster mechanism \cite{Foerster}, efficiently transfers the excitation energy to the activator, which de-excites and emits near-ultraviolet (UV) photons with a minimal emission delay. A second dopant, called shifter, is usually added to change the wavelength of the light to avoid absorption in the material. More details about the scintillation mechanism in organic materials can be found in \cite{birk1,BIRKS196439}. 
In HEP experiments, the light produced in PS is frequently collected using wavelength-shifting (WLS) fibers. These fibers shift the light from the violet/blue spectrum, which is the typical emission range of PS, to green, where the attenuation length is longer. 
Then, they guide it towards photodetectors, taking advantage of the fiber's long attenuation length, which spans over several meters \cite{Amaudruz:2012esa}.

Traditionally, PSs are manufactured with:
(1) cast polymerization \cite{cast-polymerization}, where a heated liquid monomer with dissolved dopants is poured into a mold, that results in a solid plastic structure 
after cooling;
(2) injection molding \cite{injection-molding-1,injection-molding-2}, 
where polymer granules compounded with dopants are melted and mechanically forced into a mold to solidify;
(3) extrusion \cite{extrusion-1,extrusion-2}, which pushes melted plastic through a die of the desired cross section. 
While cast polymerization provides the best optical properties, the more commonly used injection molding and extrusion offer a simpler, faster and cheaper production, hence optimal for large-volume detectors \cite{extrusion-2}.

PS detectors composed of scintillating voxels, optically isolated from each other, are optimal for particle tracking.
The typical light reflector is composed of a white diffuser, such as paint containing titanium dioxide ($\text{TiO}_2$), white polytetrafluoroethylene (PTFE), or is created using a process of chemical etching \cite{Kudenko:2001qj}.
Co-extrusion of PS and optical reflector 
has also been conducted
\cite{MINOS:2008hdf,Amaudruz:2012esa}.
Another option with good performance consists of wrapping each scintillating element with Tyvek \cite{SoLid:2018lkh}. However, such a solution is unfeasible for detectors with a large number of PS elements \cite{Sgalaberna:2017khy,Mineev:2018ekk,sfgd-testbeam-cern}.

In recent years, there have been advancements in the development of novel three-dimensional (3D) granular scintillating detectors for imaging electromagnetic and hadronic showers \cite{calice}, as well as neutrino interactions \cite{SoLid:2017ema,Sgalaberna:2017khy, Mineev:2018ekk,sfgd-testbeam-cern}.
%
Combining these geometries with the PS sub-ns response is ideal for efficient neutron detection with kinetic energy reconstruction by time-of-flight \cite{Munteanu:2019llq,instruments5040031,Gwon:2022bix,Agarwal:2022kiv}.
For instance, a neutrino detector made of two-million PS cubes, each of size 1~cm$^3$, with a total active mass of two tonnes, has been proposed \cite{Sgalaberna:2017khy}, prototyped \cite{,Mineev:2018ekk,sfgd-testbeam-cern}, 
built and 
started collecting data
as of 2023 at the T2K neutrino experiment in Japan \cite{ND280upgrade-tdr,nd280upgrade-press-release}.
Simulation studies conducted independently on a comparable 3D granularity plastic scintillator detector demonstrated that a tracking resolution ranging between 2 mm and 4 mm, depending on the utilized reconstruction algorithm, can be attained for minimum ionizing particles such as muons \cite{Alonso-Monsalve:2022zlm}.
The complexity in geometry of such a detector requires multiple vastly different 
manufacturing steps including the fabrication of every single PS cube, optical isolation via chemical etching and drilling of the holes for WLS fiber placement. 
Additionally, the assembly of two-million cubes demands a significant amount of effort \cite{Fedotov:2021ylh} and needs to be combined with a robust box
that mechanically maintains the structural integrity of the entire construction \cite{ND280upgrade-tdr}.

Attempts to simplify detector fabrication were reported in Ref.~\cite{Boyarintsev:2021uyw}, where a prototype of PS cubes glued together was obtained with a tolerance of about 200~$\mu$m. However, such a method 
is not feasible for the production of a single 3D volume of PS cubes, but only of 2D layers,   
whose production would also be time consuming for large-scale applications.
%
%
%
The above considerations call for the development of
a novel
manufacturing process that allows for
the easy production of several thousand optically-isolated PS cubes in a single block of plastic.
Our solution takes advantage of
additive manufacturing (AM), which opens the door to new automated processes that could drastically simplify the construction of future particle detectors.
More commonly known as rapid prototyping or 3D printing, AM processes build the designed parts through a layer-by-layer addition of new material. AM technologies are capable of fabricating customizable, monolithic parts with multiple materials and complex internal geometries, and can greatly reduce production time and cost.
%
%
%
Two common AM technologies for polymers are:
stereolithography (SLA) \cite{NEWMAN2015467}, where liquid resin is solidified after curing with, for example, ultraviolet (UV) light; 
and Fused Deposition Modeling (FDM) \cite{fdm}, where, analogous to extrusion, a thermoplastic material in the form of a thin wire is passed through a feeding into a melting system and deposited on a print bed line-by-line with the help of a nozzle tip \cite{astm-am-definition,fdm-description}.
3D printing of PS with SLA has been reported in literature.
However, studies have shown the necessity of either developing a new chemical composition or binding PS granules into a polymer matrix. This poses a challenge in achieving competitive performance levels in terms of acceptable light yield and attenuation length compared to standard PS \cite{3dprinted-scint-first}.
In recent years, significant progress has been made in the development of curable resins, primarily targeting applications beyond HEP. These advancements have demonstrated good performance in terms of light output, pulse shape discrimination, thermal neutron sensitivity, and various other properties when compared to commercially available alternatives 
\cite{KIM2023168537,CHANDLER2023103688,doi:10.1021/acsapm.2c00316,jne4010019,DOLEZAL2023168602,KIM20202910}.
Anyhow, 
it is not an easy task to achieve 3D printing of multiple materials, along with the production of hollow objects that possess smooth and consistent inner surfaces, which is needed for the insertion of WLS fibers.
%
%
%
As we have shown in Ref.~\cite{Berns:2020ehg}, a promising additive manufacturing option is FDM. It allows the use of the same chemical composition as standard scintillators in the form of a filament, ensures high transparency and multi-material printing.
For more details on the FDM process and the naming convention of its components, we refer to Sec.~\ref{sec:detector-fabrication}.
Furthermore, we have demonstrated that FDM can successfully 3D print \polystyrene-based scintillators. 
%
Another independent work 
reached a similar conclusion \cite{3dprinted-scint-first-fdm}. Later, we showed a technical attenuation length of about 19~cm measured in a bar of 5~cm in length, sufficient for fine-granularity scintillator detectors \cite{3DET:2022dkw}. 

The absorption and emission spectra of the 3D printed PS 
are shown in Fig.~\ref{fig:scintillation-process}.
For such a scintillator composition, the emission spectrum of the 3D printed PS peaks around 420~nm upon the excitation with 386~nm light. This result is consistent with the UPS-923A standard scintillator reported in \cite{Artikov:2005mg,Senchishin:2006qw}. 
It also shows that the measured scintillation spectrum closely matches the absorption spectrum of green WLS fibers such as the Kuraray Y11 \cite{kuraray-catalogue-wls}, allowing for an efficient readout of the scintillation light. 
In Ref.~\cite{3DET:2022dkw}, the first 3D printed matrix of PS cubes optically isolated with a custom-fabricated white reflective filament (PST, or polymethyl methacrylate (PMMA)) mixed with $\text{TiO}_2$ was produced and showed a low cube-to-cube light crosstalk, less than 2\%. 
However, after 3D printing, no work reported in literature could avoid the use of subtractive processes, like polishing the outer surface to achieve a good geometrical tolerance. Moreover, none of the studies ever attempted to 3D print hollow PS objects to host WLS fibers. 
%

In this work, we present the first-ever demonstration of additive manufacturing of a 3D-segmented, fine-granularity 
PS detector without requiring any post-processing.
%
A \supercube, consisting of a $5 \times 5 \times 5$ matrix of 1~$\text{cm}^3$ optically-isolated cubes, was additively manufactured and tested. Each cube includes cylindrical holes to house WLS fibers throughout the entire detector.

\section{Results}
\label{sec:results}

Previous R\&D on this topic carried out by the authors in \cite{Berns:2020ehg,3DET:2022dkw} did not yield a particle detector of sufficient quality, as described in Sec.~\ref{sec:introduction}. 
However, it was useful in defining the extrusion temperature of the PS at which there is no loss of the original scintillation light yield and it showed the producibility of an excellent performing optically reflective filament.

%
In this work, 
a novel manufacturing method named Fused Injection Modeling (FIM) was developed to obtain good geometrical tolerances, high transparency PS volumes, as well as precise hole fabrication for the placement of WLS fibers at a rapid production speed, thereby overcoming the aforementioned shortcomings of the AM fabricated PS particle detector. FIM merges the geometrical freedom of manufacturing of FDM with the production speed and high part density of injection molding by 3D printing an optically reflective frame containing the desired voxel shape and quantity, which is filled by rapidly pouring melted PS 
into the empty cavities to accurately shape the geometry of the PS (detailed depiction in Fig.~\ref{fig:FIM-step-by-step}).

To achieve a fast, high-quality, consistent forming of the PS structure, a customized liquefaction system integrated into an FDM machine was developed. Standard FDM melting components were analyzed and modified to cope with the increased thermal demands of a rapid heat transfer towards the low thermally conductive PS, while guaranteeing a working temperature in the filament feeding parts of the extrusion system. To distribute scintillation material evenly throughout the entire volume, an elongated nozzle was manufactured that provided freedom of movement within the already fabricated cavity. It was coupled with a spring-pressurized plate that constrained the melt pool within its mold while allowing air to escape the volume during the forming process. A more in depth explanation of the custom-made extrusion system can be found in Sec.~\ref{sec:methods}.

A Computational Fluid Dynamics (CFD) analysis was performed to determine the material requirements of the melting components, heat block and nozzle; the heat shielding parts, feeding tube and heat break; the process parameters, heat block temperature, extrusion speed, and polymer temperature at the nozzle orifice. A detailed explanation of the nomenclature of the machine components, design and mode of operation of the manufacturing process is found in Sec.~\ref{sec:detector-fabrication}.

\begin{figure*}[htb]
\includegraphics[width=1\linewidth]{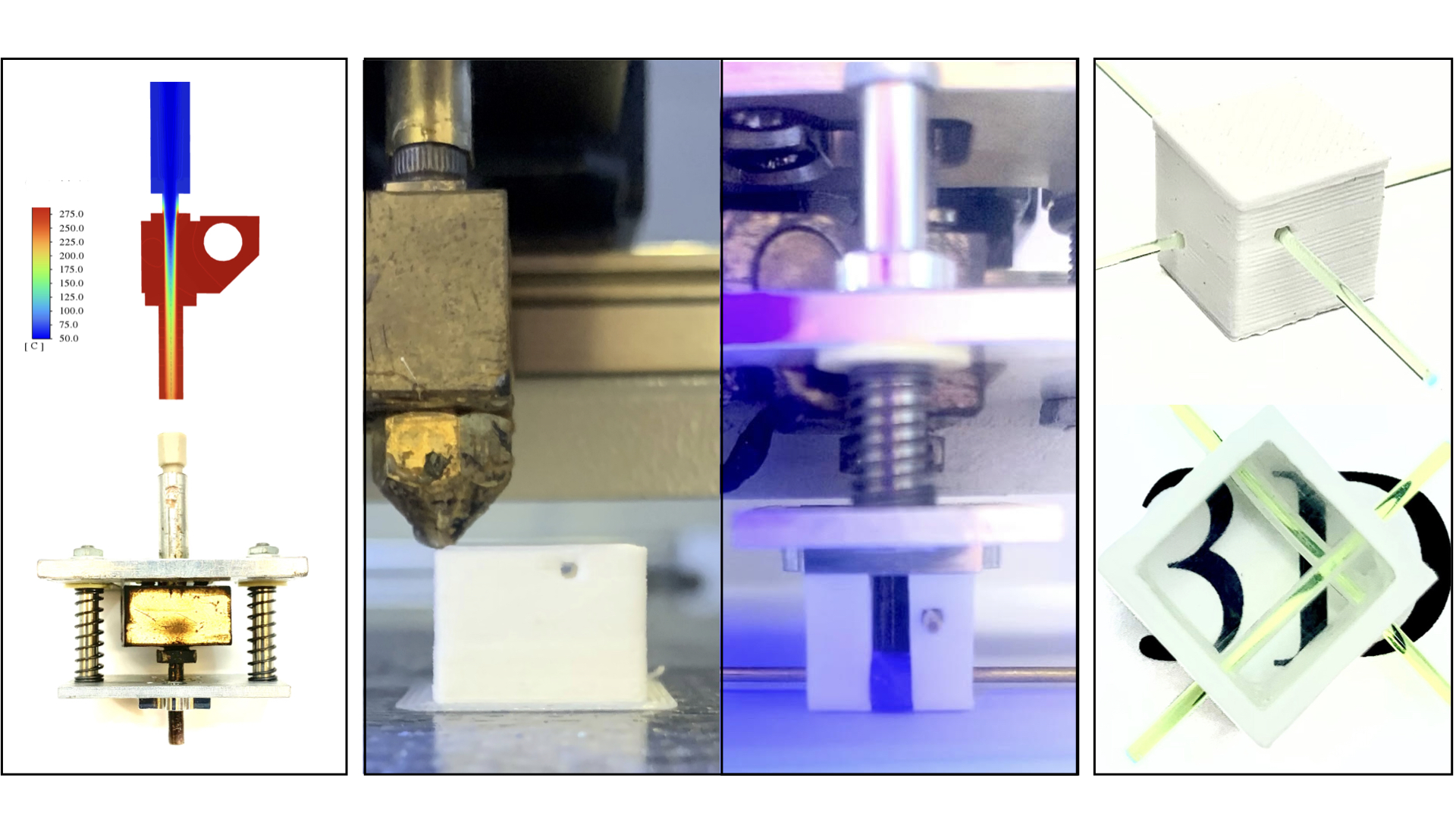}
\caption{ 
Left: (top) Result of a CFD simulation showing the temperature distribution throughout the melting system at a heat block temperature of 300°C and an extrusion speed of 15~mm/s. Components from top down: filament feeding tube (blue), heat break (multi-colored), heat block (red), nozzle (red), and PS filament (multi-colored, flowing through the whole system). Combined with the chosen component materials, these parameters resulted in a PS temperature at orifice of around 230°C (orange color at nozzle tip) and a steep temperature reduction upwards to protect against clogging issues (shift from red to blue at the heat break).
(bottom) Custom melting system for PS forming. Elongated copper nozzle enables movement to the bottom of the already manufactured reflective cavity. Two springs allowing the nozzle to slide through the pressure plate while also constraining the melt pool at the end of the filling procedure.
Middle: 
Actual demonstration of the FIM process. 
First, the reflective walls are 3D printed to obtain the optically reflective mold (middle-left), then the empty cavity with inserted metal rods that create space for WLS fibers is filled with melted PS (middle-right). For a better illustration, the reflective wall has an open slit in the front to show the flow of the melted PS from the extruder into the cavity, while being illuminated with UV light. The conceptual steps of the FIM process are shown in Fig.~\ref{fig:FIM-step-by-step}.
Right: a PS cube manufactured with FIM and instrumented with WLS fibers. The PS cube is shown both covered by white reflector on all its six faces (top) and open on the bottom and top to show the inner PS transparency of the filled volume and the retained squared shape of the reflective mold after filling had occurred (bottom). The bottom face was polished to remove the wall roughness created by the pattern of the FDM produced frame during filling.
\label{fig:FIM-method}  
}
\end{figure*}

The results of the CFD analysis concluded that both the heat block and the nozzle needed to be manufactured with copper. Its high thermal conductive property enables a rapid heat transfer to the low thermally conductive \polystyrene-based PS filament, such that the desired melt temperature could be reached at high extrusion speeds. The feeding tube had to be fabricated with polyether ether ketone (PEEK), the heat break with stainless steel and a wall-thickness of 250~$\mu$m. The low thermal conductivity of these materials combined with the small cross-section of the heat break results in a highly thermal resistant structure that prohibits heat flow towards the temperature sensitive parts and maintains a working temperature of the whole extrusion system. This composition of components resulted in the optimal process parameters of: a maximum throughput-speed of 15~mm/s to generate a high mass flow that quickly spreads within the cavity before solidifying; at a peak heat block temperature of 300°C, with which premature melting followed by system clogging can be avoided. This combination results in a temperature of around 230°C throughout the entire cross-section of the PS at the nozzle orifice, which preserves the scintillation properties established in \cite{Berns:2020ehg,3DET:2022dkw} (Fig.~\ref{fig:FIM-method}, left). 

The performance of the custom heat break was tested by measuring its temperature with a thermocouple at a heat block temperature of 300°C and no filament extrusion. The results showed a 43.2\% temperature reduction from 155°C to 88°C compared to a standard model heat break. This value lies below the glass transition temperature of the PS (100°C), thus ensures a failure free extrusion process.

The fabrication procedure, using the custom manufactured modified extrusion system (Fig.~\ref{fig:FIM-method}, left, bottom), contained three main steps and is described in more detail in Sec.~\ref{sec:methods}: the fabrication of one matrix-layer of the reflective frame via FDM; the insertion of metal rods into the voxel cavity through already 3D-printed holes in the frame; followed by the forming of the cube-shaped volume with melted scintillation material (Fig.~\ref{fig:FIM-method}, middle). These processes can be repeated until a \supercube~of the desired size has been obtained. Finally, the WLS fibers were placed through the PS voxel via the cylindrical holes produced by the removed metal rods (Fig.~\ref{fig:FIM-method}, top right).
Metal rods were essential because WLS fibers could not be positioned during the PS forming process without risking thermal damage.

The optically reflective frame constituted the mold with which the PS was formed and thus needed to withstand the pressure and heat from the extruded melt pool.
As reported in our previous work~\cite{3DET:2022dkw}, in order to manufacture a 3D printed layer of optically-isolated cubes, a custom optical reflector filament was fabricated with either PMMA or PST. 
Although a transmittance of less than 10\% and an optical cube-to-cube light crosstalk of less than 2\% were achieved,
the material could not maintain its structural integrity during the injection of the PS. This was due to the similar heat resistance of the optical reflector compared to the PS, with the consequence of swelling and wall bending during filling. Hence, for the FIM method, a white, more heat-resistant polycarbonate-PTFE filament was used, which combines good optical properties of PTFE and a heat-resistance significantly higher than the previously used materials. 

In FDM, vertical walls are created by stacking lines on top of each other, while horizontal walls are formed by placing adjacent lines on the same print plane.
This results in different fill factors, thus different light transmissive properties. The corresponding transmittance for a wavelength of 420~nm, measured with a monochromatic light source in air, resulted in 13\% for horizontally and 18\% for vertically built walls, as shown in Fig.~\ref{fig:transmittance}. Consequently, to obtain a uniform cube-to-cube light crosstalk, horizontal walls were fabricated with a thickness of 1.2~mm and the more transmissive vertical walls with 1.5~mm. The more heat-resistant reflective material retained its as-built cube shape throughout the filling fabrication of the PS, as can be seen in Fig.~\ref{fig:FIM-method}, bottom right. Caliper measurements of vertical walls showed an average thickness of $t_{\text{walls, mean}}$ = 1.51~mm, deviating by only 0.01~mm from the nominal $t_{\text{walls, designed}}$ = 1.5~mm with a standard deviation of $t_{\text{wall, std}}$ = 0.01~mm. Both the retention of the reflective frame geometry after filling and the accurate fabrication of the wall thickness ensured a consistent active volume of scintillation material in every voxel. More details about the fabrication of the white reflector walls can be found in Sec.~\ref{sec:methods-white-reflector}.

The established process parameters combined with the custom manufactured elongated nozzle and melt-pool-containing pressurized plate resulted in a transparent PS volume with a high fill factor around the metal rods, which left precisely positioned and close-fitting holes for the insertion of WLS fibers. In Ref.~\cite{Berns:2020ehg} a technical attenuation length of about 19~cm was measured and found to be sufficient for highly-segmented detectors. The FIM-fabricated PS showed an even improved transparency and no air bubbles compared to the samples produced with FDM (Fig.~\ref{fig:FIM-method}, bottom right).


%
%

The manufacturing time per an entire unit, depicted top right in Fig.~\ref{fig:FIM-method}, was approximately 6 minutes. This involved the fabrication of the reflective frame and the forming of the PS. Fixed time cost, including the 
not-yet fully automated change between the FDM and injection set-up, was around 20 minutes, plus an additional 15~minutes for an initial warm-up of the machine.

\begin{figure*}[htb]
   \includegraphics[width=0.49\linewidth]{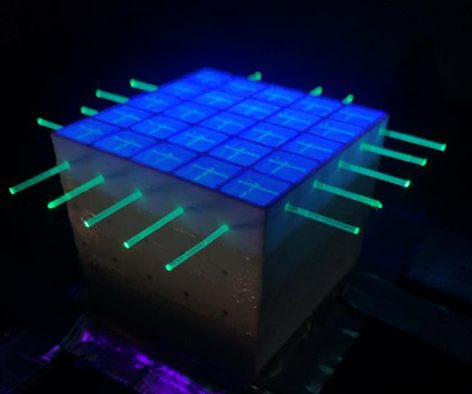}
   \includegraphics[width=0.49\linewidth]{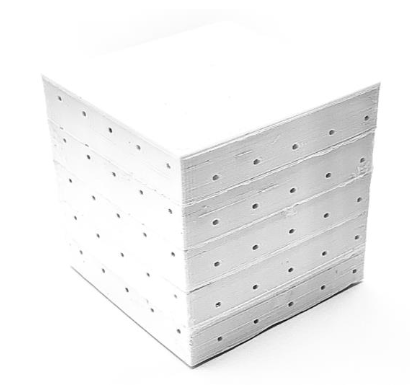}
\caption{ 
Left: \supercube~illuminated with UV light prior to completion. 
The absence of a white reflector layer on the \supercube's top surface provides a clear view of the inner structure, featuring optically isolated PS cubes intersected by WLS fibers.
Right: FIM manufactured 5x5x5 \supercube, requiring no post-processing. 
\label{fig:SuperCube}  
}
\end{figure*}

\begin{figure*}[htb]
   \includegraphics[width=0.45\linewidth]{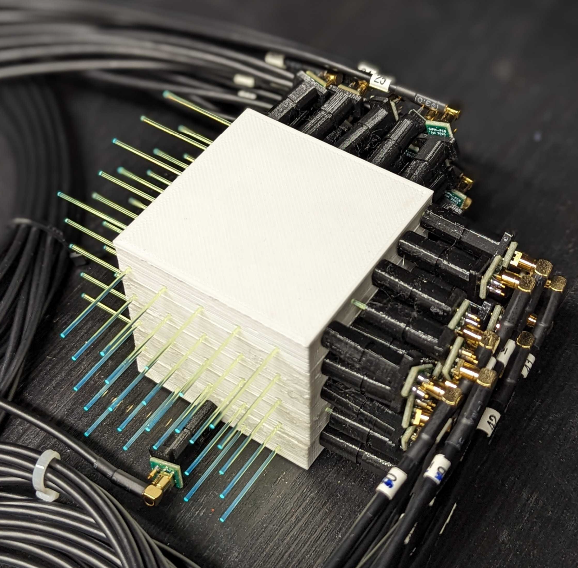} \\
   \includegraphics[width=0.49\linewidth]{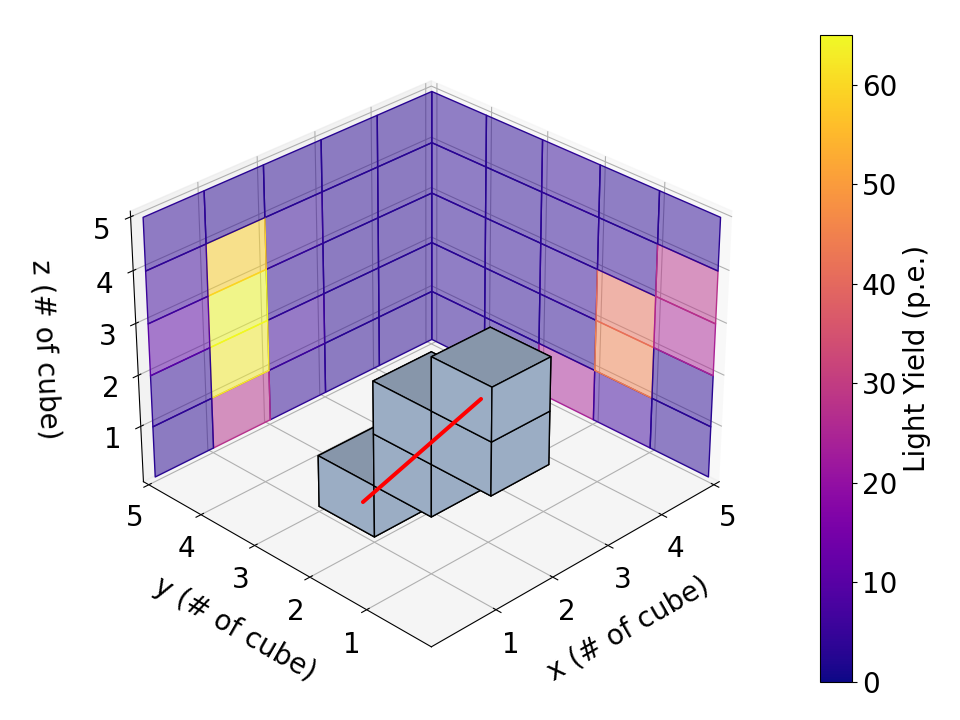}
   \includegraphics[width=0.49\linewidth]{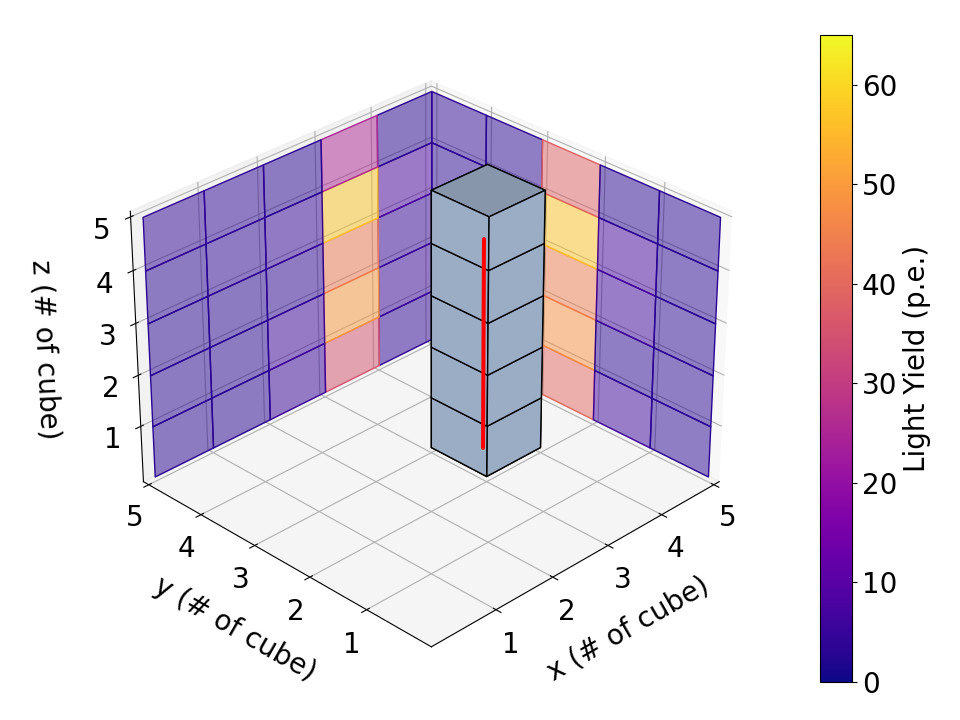}
\caption{ 
Top: FIM manufactured \supercube~prototype instrumented with WLS fibers and SiPMs.
Bottom: events of two detected cosmic particles crossing the \supercube~from top to bottom with a detected diagonal (left) and vertical (right) track. The 2D projections show the number of photons detected in each readout channel (color scale), while the 3D voxels show the reconstructed 3D particle track. In the matching of the 2D projections channels with less then 7 PE, likely from cube-to-cube light crosstalk, were not considered. The red line indicates the fitted particle trajectory.
\label{fig:SuperCube–cosmics-event-display}  
}
\end{figure*}

The $5 \times 5 \times 5$ voxel \supercube, obtained using the FIM method without the need of post-processing or additional steps, is shown in Fig.~\ref{fig:SuperCube}.  It was instrumented with WLS fibers that capture and guide the scintillation light produced by charged particles in single cubes towards coupled silicon photomultipliers (SiPM), which count the number of photons impinging on their active surface. 
The instrumented detector is shown in Fig.~\ref{fig:SuperCube–cosmics-event-display}.
More details can be found in Sec.~\ref{sec:detector-cosmics}.

\begin{figure*}[htb]
   \includegraphics[width=0.49\linewidth]{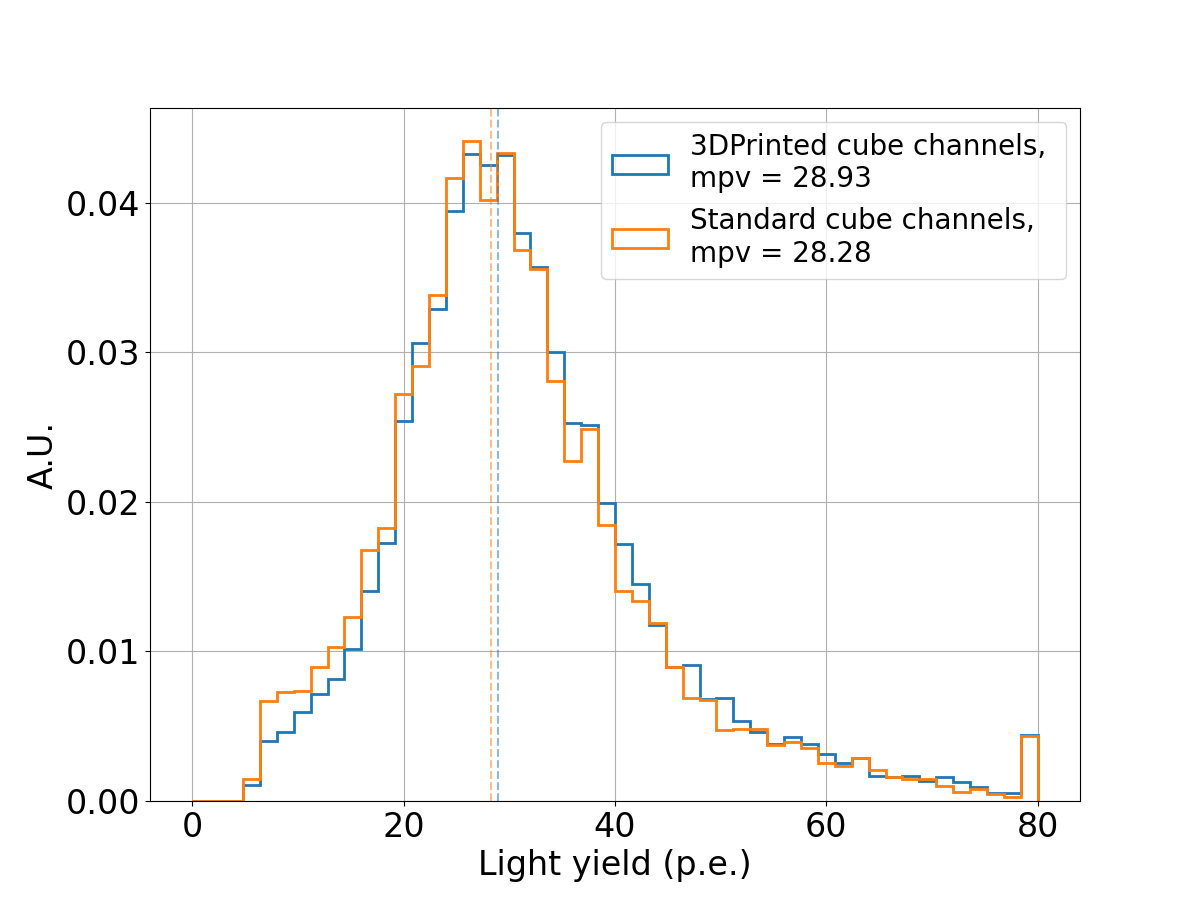}
    \includegraphics[width=0.49\linewidth]{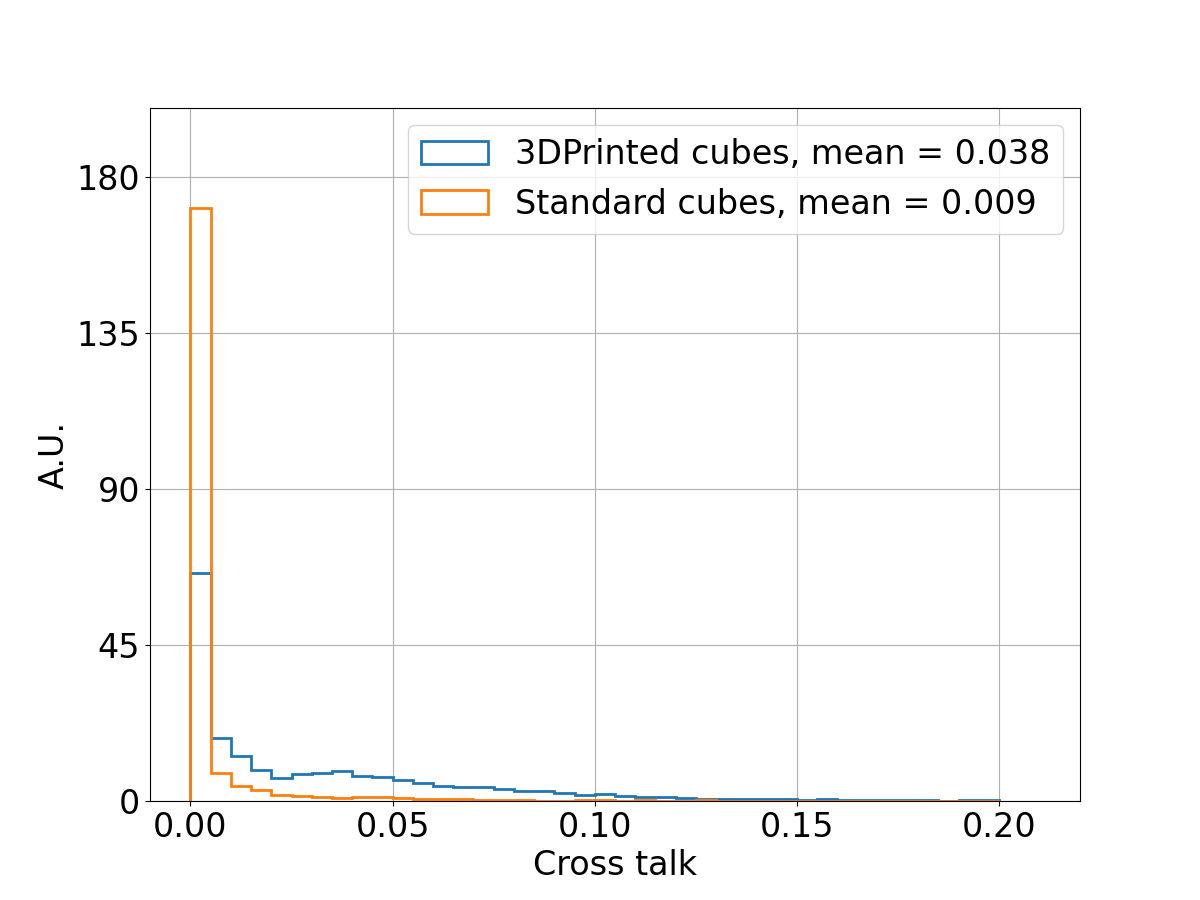}
\caption{ 
The light yield (left) and the cube-to-cube crosstalk (right) measured from down-going cosmic particles crossing the \supercube~3D produced with FIM (blue) and a prototype produced with 
cast polymerization
(orange) is shown.
The vertical dashed lines indicate the MPV of the measured light yield distributions. 
\label{fig:SuperCube–cosmics-results}  
}
\end{figure*}

The response of the \supercube~was characterized with cosmic particles in terms of single-cube scintillation light yield and cube-to-cube scintillation light crosstalk, which are the main parameters that determine the detector performance in terms of particle tracking, identification and calorimetry. Two detected cosmic particle events are shown in Fig.~\ref{fig:SuperCube–cosmics-event-display}. Particles producing a vertical track, whose range in the detector can be more precisely estimated, were selected and used for analysis.
Being mostly minimum ionizing particles (MIP), their typical energy deposited in PS is approximately 1.8~MeV/cm, which allows to precisely estimate the number of photons detected per unit energy loss.
Moreover, with an energy spectrum that spans from a few hundred MeV to several GeV, cosmic particles typically produce through-going tracks in the detector with a very uniform energy deposition. 
By measuring the number of PE across 
all the readout channels and geometrically matching the two detector  
projections of the event, the 3D track of the particle can be 
reconstructed and the number of scintillation photons produced in each cube can be accurately measured.
As shown in Fig.~\ref{fig:SuperCube–cosmics-results}, the scintillation light yield of the 3D printed prototype was found to be comparable to that obtained with the one produced by 
cast polymerization
(using the same scintillator composition) in \cite{Boyarintsev:2021uyw}.
A most probable value (MPV) of about 29 PE was measured for the 3D printed \supercube, close to the MPV of the prototype produced with 
cast polymerization
(about 28 PE). 

In conclusion, the newly developed FIM method allowed for the production of a monolithic 3D optically-segmented $5 \times 5 \times 5$ matrix of scintillating cubes measuring each $1~\text{cm}^3$. These cubes were fabricated with accurately placed holes for the insertion of WLS fibers in both the X and Y directions.
No post-processing was necessary and the detector could be instrumented with the read out electronics right after the fabrication process was completed.
Hence, for the first time, a PS detector capable to track elementary particles and measure their energy loss has been successfully additively manufactured.
Experimental tests showed a detection performance comparable to the state of the art detectors produced with traditional manufacturing techniques.

\section{DISCUSSION}
\label{sec:discussion}

The innovative fabrication process developed recently has made it possible to capture a 3D image of particle interaction using a particle detector that was entirely produced through additive manufacturing for the first time.
The development of the FIM method was essential to achieve the improved results described above, as it brought several advantages over FDM and other more standard methods like injection molding 
or cast polymerization.

The FDM manufactured PS showed a sufficient transparency for centimeter-sized voxels, but the FIM fabricated samples visibly improved the quality by reducing imperfections created by the merger of different deposited lines during the FDM process and the reduction of air bubbles within the PS volume. 
The improved transparency results from the straight down-to-up filling pattern, which creates a single melt pool that expands and forms the PS structure within the optical reflective frame. 
The nozzle's orifice remained submerged in the melt pool until the PS forming process was complete.
This technique prevented air from being trapped between different layers of melted polymer.
Additionally, the forming of a cubic-centimeter sized PS cube with FIM needs around 30~seconds, 
whereas the same structure takes around 5~minutes using FDM fabrication, i.e., about ten times as long. The faster per-unit manufacturing time is crucial when it comes to the production of large volume, highly-segmented particle detectors that can include millions of isolated voxels.

The optical isolation of PS voxels is ensured by the FIM process via the FDM-manufactured reflective frame. This is superior to 
traditional methods,
which requires additional vastly dissimilar fabrication steps,
as described in Sec.~\ref{sec:introduction}.
The sequence of different fabrication processes increases the total time and cost of production due to the involvement of multiple manufacturing parties, transportation and dead time between steps. 

The geometrical tolerance of the optically isolating frame produced by FIM is improved compared to the particle detector manufactured solely by FDM. This is a result of the introduction of a heat-resistant material that is able to preserve its designed shape during the filling of melted PS, thus creates a distinct boundary between the individual PS voxels. In FDM, both the scintillation and the reflective filament are made of plastics with similar thermal characteristics, which can result in a mixing at the boundary layers between these two materials and also a warping and distortion of the reflective geometry. Although the tolerance cannot be considered as good as the one achievable with injection molding 
or cast polymerization
(better than $50~\mu\text{m}$ for plastics), it is worth noting that it is affected by the type of 3D printed material, the printing strategy and the precision of the machine, which will be improved in future studies.

The largest advantage of the FIM method compared to FDM 
or more traditional methods
is the capability of creating precise, close-fitting holes through the entire particle detector to host WLS fibers. FDM is notorious for producing poor quality small holes in terms of size consistency, shape and surface roughness, which results in a low WLS fiber to PS contact area, especially perpendicularly to the print plane. 
Injection molding has the capability to create hollow components within the manufactured structure. However, incorporating a white reflector without causing any harm to the pre-existing holes poses a significant challenge.
In our next prototypes also a third vertical hole will be added.
An alternative option would be to utilize thin pipes capable of withstanding temperatures significantly higher than 200°C, in place of metal rods. This modification would facilitate the easy insertion of WLS fibers, leading to the creation of the SuperCube. A single-layer prototype was successfully produced using borosilicate glass pipes with outer and inner diameters of 1.5 mm and 1 mm, respectively. While this approach eliminates the need to extract rods after the injection and solidification of the plastic scintillator, the addition of non-plastic material does slightly increase the inactive volume. Consequently, we opted to prioritize the baseline option involving metal rods.

In terms of the detection performance, the quantity of scintillation light, collected from the FIM manufactured PS cubes crossed by cosmic particles, is similar to that observed from PS cubes produced 
with traditional methods like cast polymerization.
The cube-to-cube crosstalk at the few-percent level is comparable to that of typical PS detectors \cite{sfgd-testbeam-cern} and is low enough to provide unambiguous 3D particle tracking.

The light crosstalk is about four times higher than that with the 
cast polymerization
prototype, 
due to the higher wall transmittance. Hence, one would expect an approximately 20\% lower total light yield. However, this was not observed, as shown in Fig.~\ref{fig:SuperCube–cosmics-results}.
One may conclude that the additive manufactured PS has an intrinsic light output higher than the one from 
cast polymerization,
but this was not the result of our past work \cite{3DET:2022dkw,Berns:2020ehg}.
It is well known that the light yield of a PS detector can be enhanced by improving the light trapping efficiency of the WLS fibers through the reduction of air gaps, that is the increase of the contact area between the PS and the WLS fibers, whose refractive indices are very similar and thus allow a high transmittance from the scintillation material to the fiber. This can be obtained by using optical grease \cite{Artikov:2016sqg,Artikov:2017fmr,Artikov:2016uem}, which is difficult to homogeneously introduce in long, small-diameter holes. Using the FIM method, it was possible to create holes, that span the length of the \supercube,~of only 100~$\mu$m larger than the WLS fibers, resulting in a large contact surface between the PS cubes and the fibers. This feature is expected to increase the total light yield and to compensate for the light loss from the observed crosstalk. 



A preliminary time resolution measurement was conducted using a Hamamatsu H6410 photomultiplier tube (PMT) coupled to a sample of 3D printed plastic scintillator exposed to a $^{60}\text{Co}$ radioactive source. The data was read out with a LeCroy Wave runner 104MXi-A oscilloscope, revealing a decay time consistent with that of UPS-923A, indicating no degradation of the timing properties resulting from the scintillator additive manufacturing process. However, a more comprehensive quantitative characterization is deferred to future work and publication.

Aging presents another critical concern, potentially resulting in a few percent reduction in light yield annually for extruded polystyrene-based scintillators \cite{Artikov:2005mg,T2K:2022atj,MICHAEL2008190,ALIAGA2014130}. Given the similarity of the FIM process to extrusion or injection molding in the treatment of the plastic scintillator material, a similar degradation is expected. Despite various measurements conducted over approximately four months, no visible changes in light yield and crosstalk were observed within the precision of the experimental setup. Nevertheless, more precise quantitative results require dedicated aging measurements, which are planned for future investigation.

Overall, the FIM method results in a faster particle detector fabrication by including the shaping of the PS in optically-isolated voxels into an already assembled final detector using one single machine.
From here, we see no obstacles that would hinder the viability of this method for large scale particle detector production. Each manufacturing step has consistently yielded satisfactory results. 


%
In conclusion, we have successfully achieved the additive manufacturing of a 3D optically-segmented plastic scintillator detector. To the best of our knowledge, this is the first time such a feat has been accomplished. This detector is capable of both particle tracking and energy loss measurement.
It demonstrates a tool that opens new possible concepts of particle detectors characterized by a very fine 3D granularity in very large active volumes, above the tonne-scale.
Although not tested in this work, millimeter-sized granularity and bigger units, e.g. $30 \times 30 \times 30~\text{cm}^3$ or above, are expected to be achieved without problems.
We believe that this new technology opens up new possibilities for the future of particle physics experiments.
In particular, long-baseline accelerator neutrino oscillations 
\cite{Abe:2017uxa,Abe:2019vii,NOvA:2021nfi,Abe:2018uyc} 
or those searching for new neutrino sterile states via short-baseline oscillations 
\cite{SoLid:2020cen,PROSPECT:2020sxr,STEREO:2019ztb,RENO:2020hva,DANSS:2018fnn,Serebrov:2020kmd} 
can profit of the developed additive manufacturing process by building very large but finely-segmented scintillator detectors to obtain a large sample of very detailed images of neutrino interactions.
Moreover, the developed production method will allow to easily build highly-performing fine-granularity electromagnetic or hadronic sampling calorimeters, enabling high-resolution particle flow analysis and fulfill the requirements of future collider experiments \cite{calice,CALICE:2018ibt}.
FIM can be performed with various types of filaments, where the white reflector could be covered by an additional layer made of heavier nuclei or doped material. Finally, large-volume  3D detector geometries, made possible by the FIM process, can allow for a high-efficiency detection of fast neutrons with a precise measurement of their time of flight, thanks to the large content of hydrogen in PS and its short decay time \cite{Sgalaberna:2017khy,Gwon:2022bix}.
This is of fundamental importance for next generation neutrino physics experiments, providing constraints on leading systematic uncertainties
\cite{instruments5040031,MINERvA:2023avz,Munteanu:2019llq,Baudis:2023tma}.


\section{METHODS}
\label{sec:methods}


\subsection{Particle Detector Geometry}

\begin{figure}[]
   \includegraphics[width=0.8\linewidth]{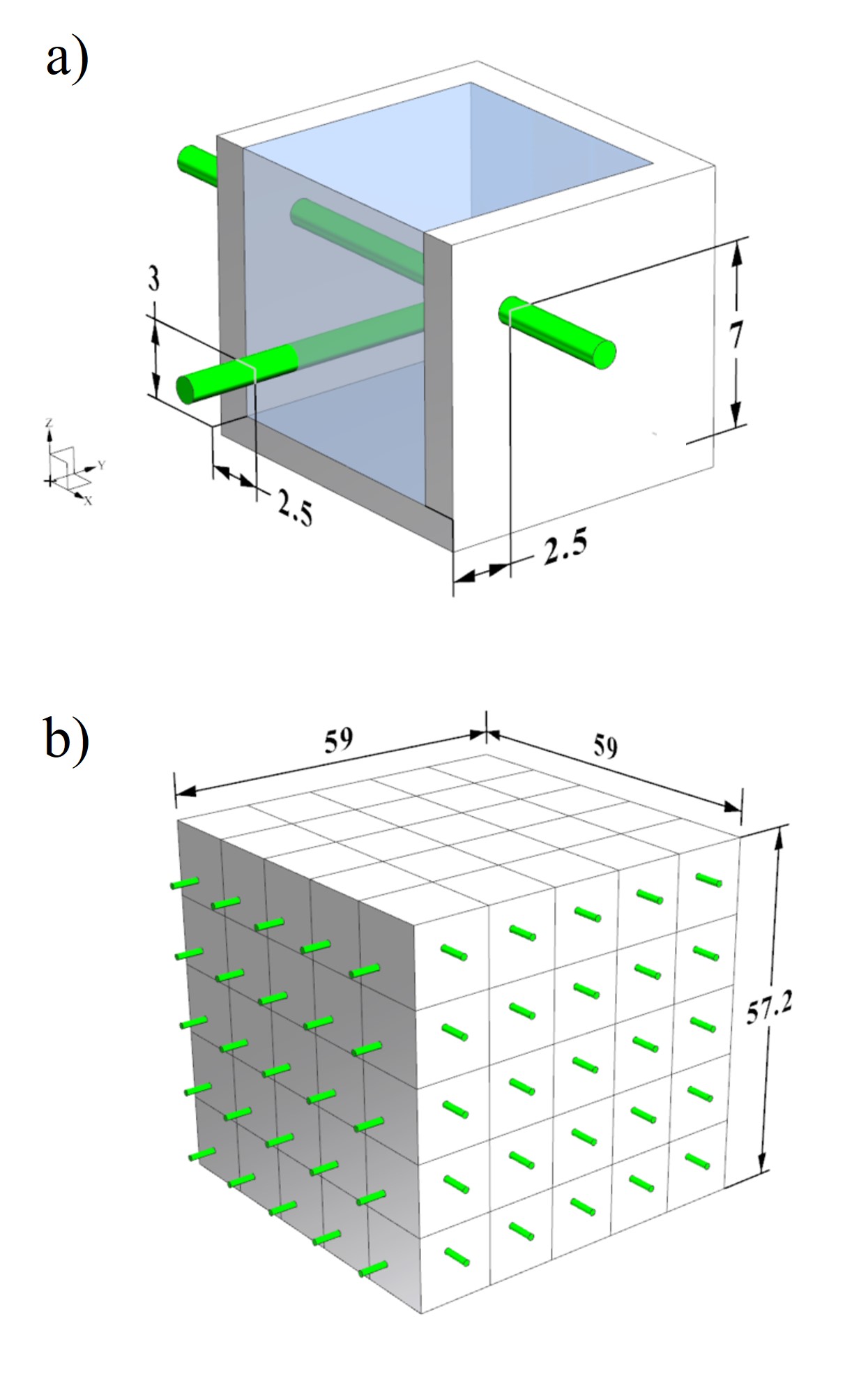}
\caption{ 
\label{fig:PD geometry}
a) Components of a voxel with parts of the reflective frame cut away on top and side. The reflective shell (white) enclosing the cube-shaped PS (blue) traversed by two wavelength shifting fibers (green) with their respective coordinates (in mm).
b) Depiction of a $5 \times 5 \times 5$ voxel \supercube~with its outer dimensions (in mm).
}
\end{figure}

The overlaying structure of particle detector is a cuboid that consists of smaller, cube-shaped detection-units called voxels. One voxel consists of three elements: a $10 \times 10 \times 10~\text{mm}^3$ PS volume for visible light production; a reflective frame with a wall thickness of 1.5~mm in vertical and 1.2~mm in horizontal plane direction, enclosing the PS to entrap the light in the unit; and two perpendicularly arranged WLS fibers in the horizontal plane with the coordinates (X = 2.5~mm, Z = 3~mm) in Y-direction and (Y = 2.5~mm, Z = 7~mm) in X-direction that absorb the light from the PS, shift its wavelength such that the event can be read-out by an external system connected to the fibers (see Fig.~\ref{fig:PD geometry} a and Sec.~\ref{sec:detector-cosmics} for details). 
The prototype of the ``\supercube'' detector consists of 125 detection units arranged in a $5 \times 5 \times 5$ configuration, resulting in total dimensions of 59 mm in width and 57.2 mm in height, as illustrated in Fig.~\ref{fig:PD geometry} b.

\subsection{Particle detector fabrication}
\label{sec:detector-fabrication}

\begin{figure}[]
   \includegraphics[width=1\linewidth]{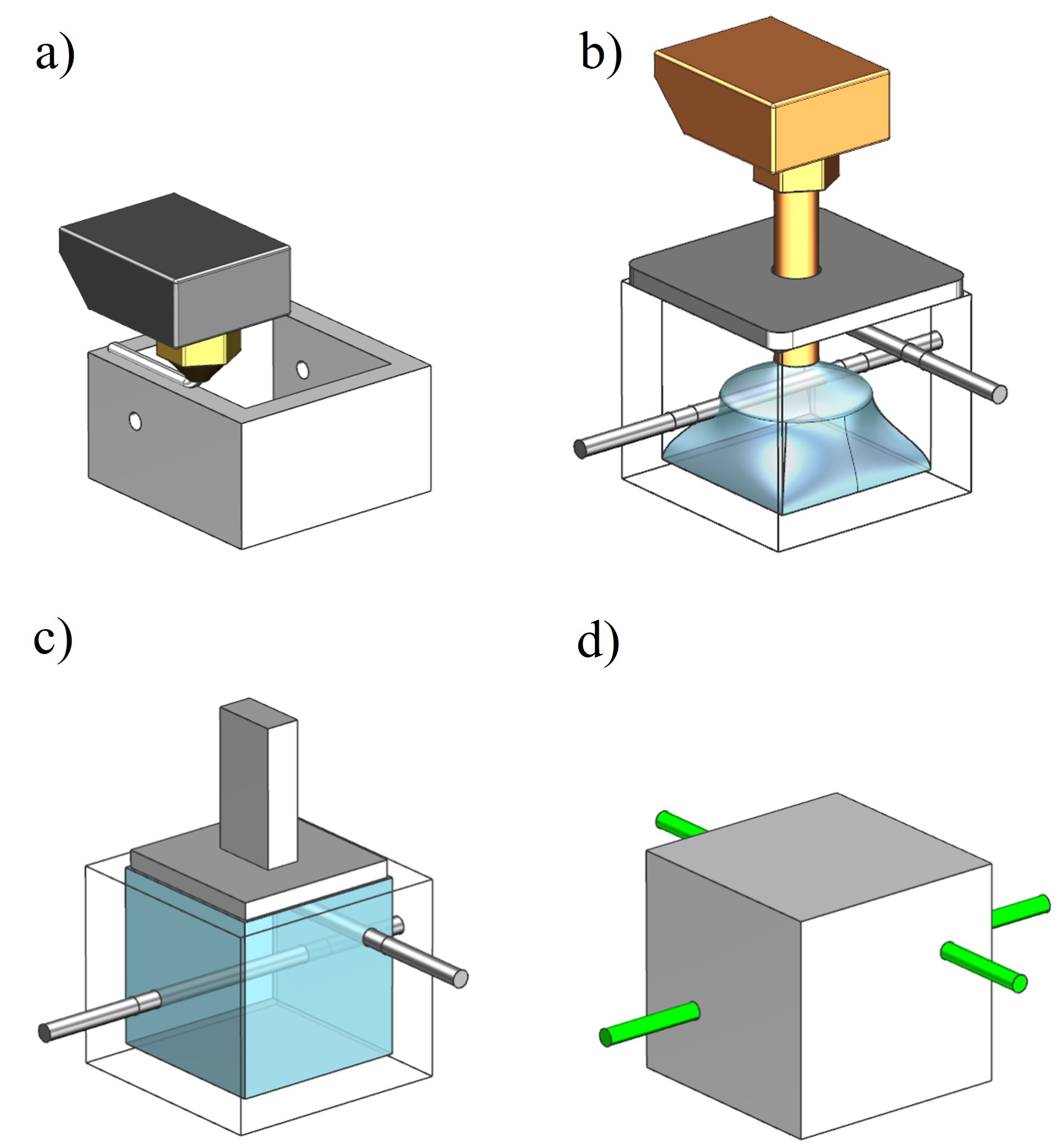}
\caption{ 
\label{fig:FIM-step-by-step}
Depiction of the fabrication process of one particle detector voxel. a) Fabrication of the reflective frame (white) via FDM. b) Filling of the scintillation material with customized extrusion components. Metal rods (black) create a tube-shaped void for the WLS fibers. Pressurized plate on top of the voxel constraints the melt pool within the cavity. c) Planing of the top surface to ensure a successful continuation by FDM. d) Finished voxel with a 10x10x10 mm scintillating cube surrounded by a reflective shell with WLS fibers (green) inserted through the whole structure.
}
\end{figure}

The \supercube~was produced in two separate processes; the fabrication of a $5 \times 5$ matrix layer of an optically reflective frame and the forming of the PS, each using a different mechanical 
setup. 
The two production steps were alternated to build a layer-by-layer three-dimensional $5 \times 5 \times 5$ \supercube. The reflective frame with holes for the WLS fibers was manufactured via FDM without the use of support structure (Fig.~\ref{fig:FIM-step-by-step} a).
 The PS was formed in three sub-processes. First, metal rods were placed through the holes to create circular voids for the WLS fibers to be inserted, then the square-shaped volume was rapidly filled in a bottom-to-top motion with a customized extrusion setup combined with a pressurized plate to keep the melt constrained in the cavity (Fig.~\ref{fig:FIM-step-by-step} b) and thirdly, a heated punch was pressed on top of the PS to plane its surface such that the next matrix-layer could be fabricated on top of it (Fig.~\ref{fig:FIM-step-by-step} c). Finally, the top of the voxel was closed by a top layer and the WLS fibers were inserted through the already manufactured holes (Fig.~\ref{fig:FIM-step-by-step} d).

\subsubsection{Reflective Frame Fabrication}
\label{sec:methods-white-reflector}

The reflective shell was manufactured via 
FDM using a \textit{F430} from \textit{Creatbot} \cite{creatbot-f430}. In FDM, a polymer string is processed through an extrusion system, which is divided into a cold and hot zone, enabling the distribution of the material with a predetermined temperature and shape. The role of the cold zone is to push the filament with the appropriate speed through feeding tubes into the hot zone, using a motor-controlled gear and roller structure. In this part of the extrusion procedure, the machine temperature must be below the glass transition temperature of the polymer-in-fabrication, otherwise premature softening of the string leads to an increase in friction with the feeding tubes, followed by clogging and system failure. The purpose of the hot zone is to rapidly melt the low thermally conductive material through a metal heat block kept at a polymer specific temperature, such that it can be distributed via an extrusion nozzle into the desired shape. The link between both zones is the heat break, a thin circular tube, which acts as a thermal resistor. It prohibits the heat from creeping from the hot into the cold zone, thus establishing a working temperature in both areas of the extrusion system.  

A 1.75~mm polycarbonate-polytetrafluoroethylene-blend (PC-PTFE) filament from \textit{Rosa3D filaments} \cite{rosa3d-filament}
was used for the frame fabrication. This material provided the required combination of good optical reflectivity to contribute to the performance of the detector and high thermal resistance to ensure the dimensional integrity of the geometry during the forming of the PS.\\
The structure was printed with a nozzle temperature of 280°C, a bed temperature of 115°C for the first matrix layer and a chamber temperature of 55°C to reduce the risk of warping. 
After the first filling cycle, the bed temperature was set to 75°C at all time to avoid permanently turning the scintillation material opaque, as was observed in previous studies \cite{3DET:2022dkw}. The reflective structure was fabricated using a 0.4 mm diameter nozzle with a fill factor of 100\% for both horizontal and vertical walls. The layer height was set to 0.2 mm, the print speed to 2000 mm/min, and cooling was activated only during the layers incorporating circular holes to ensure accurate round geometries without the need for support structures. In FDM, the thickness of horizontally-printed walls (parallel to the print bed) is very close to the designed value and was not further investigated. However, 20 thickness measurements of vertically-printed walls (perpendicular to the print bed) were taken using a caliper to show the accuracy and precision of this manufacturing step.

\begin{figure}[h]
    \includegraphics[width=.9\linewidth]{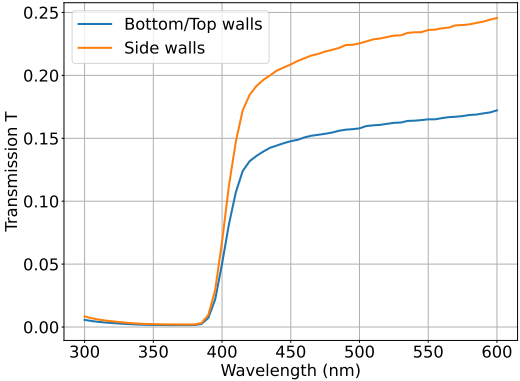} 
    \caption{ 
    Measured transmittance of white reflector sheets with the same thickness but different print orientation compared to the printing bed. Vertical (side) walls in red and horizontal (top and bottom) walls in blue of a PS voxel.  
    The experimental data were collected in air.
    \label{fig:transmittance}  
    }
\end{figure}

The measured transmission for bottom, top and side walls of the white reflective shell, taken in air, is shown in Fig.~\ref{fig:transmittance}.
Data were collected with a monochromatic light source and read out with an integrated sphere and a photodiode (Hamamatsu S1337-1010BQ).

\subsubsection{Plastic scintillator forming}
The PS fabrication consisted of three sub-objectives: maximizing the performance of the particle detector, access to an event read-out using WLS fibers in two directions (X, Y) and the preparation of the matrix-layer-in-fabrication for an additional matrix-layer to be built on top. 

To optimize the performance of the particle detector, the material required high transparency and a consistent fill factor throughout the voxel to maximize the active material volume. Additionally, it was crucial for the PS to surround the metal rods, ensuring a substantial contact area between the scintillation material and the WLS fibers to enhance light capture efficiency.
The fill density was influenced not only by the viscosity of the polymer but also by the magnitude of the mass flow rate exiting the nozzle. Increasing the flow rate resulted in a more energetic distribution of the melt within the cavity, exerting greater pressure on the material towards the boundaries. This accelerated the distribution of the melt without allowing the polymer to cool and recrystallize before reaching the outer walls of the reflective shell.
An elongated extrusion nozzle, capable of reaching the bottom of the reflective shell, facilitated an even distribution of scintillation material throughout the entire cavity. The liquid polymer was deposited through the 1.8 mm wide nozzle orifice in a vertical bottom-to-top motion, forming a single blob from which the melt spread towards the boundaries. Positioned at the top, a spring-loaded, polished stainless steel plate constrained the melt pool within the cavity. The plate incorporated a hole to enable unrestricted vertical movement of the nozzle, along with vent holes designed to release air during voxel-filling. Additionally, it featured a concave dome to provide temporary overfill, effectively compensating for any subsequent material shrinkage.

%

\begin{figure}[h]
   \includegraphics[width=1.1\linewidth]{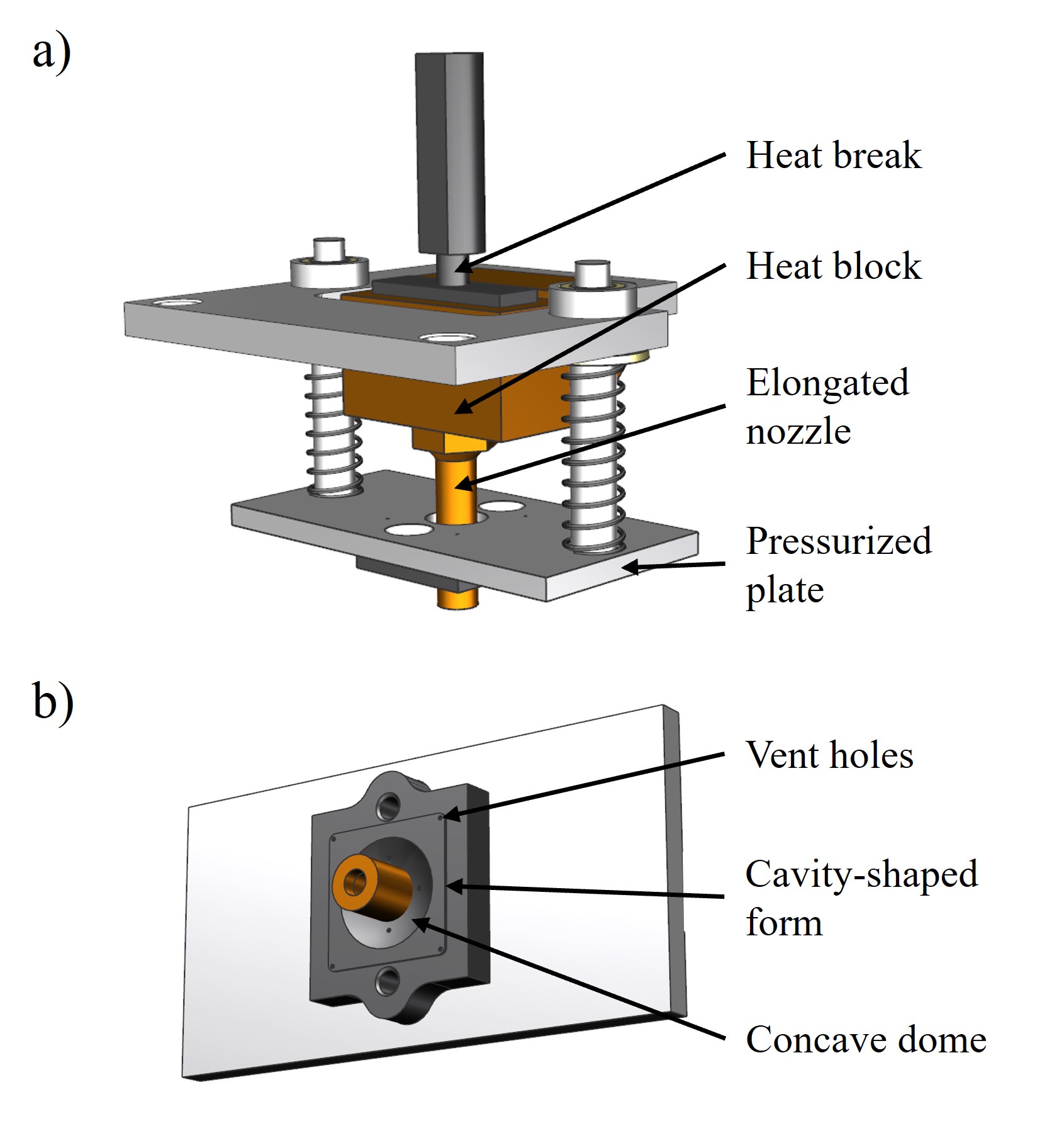}
\caption{ 
\label{fig:Extrusion system}
a) Depiction of the extrusion system components modified for the filling of scintillation material into the reflective cavity. b) Geometry of the pressurized plate that is pressed onto the cavity to keep the melt pool restrained in fill volume. 
}
\end{figure}

\begin{figure}[H]
   \includegraphics[width= 1\linewidth]{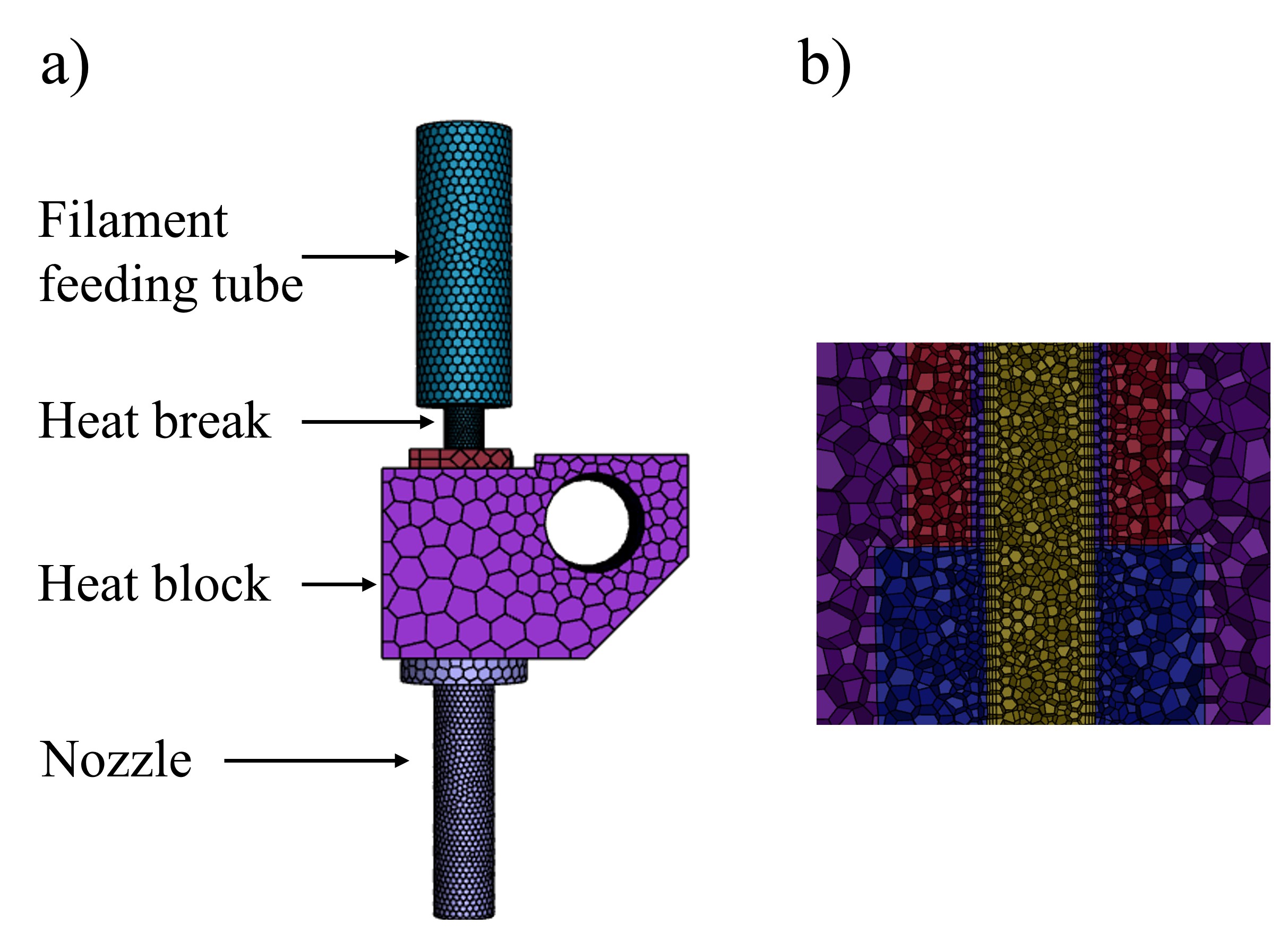}
\caption{ 
\label{fig:CFD model}
a) Components of the CFD model. The black mesh depicts the discretization points of the structure. b) Cross-section view of the inside of the model depicting the polyhedral volume mesh (purple - heat block, red - heat break, blue - nozzle and yellow - polymer). 
}
\end{figure}

A CFD analysis using ANSYS Fluent was performed to determine the material requirements of the melting components heat block and extrusion nozzle, the geometry and material of the heat break and the feeding tube of the filament, and the process parameters heat block temperature and extrusion speed (Fig.\ref{fig:CFD model} a). \\
These process elements were balanced to attain a PS outflow temperature at the nozzle orifice of around 230°C to preserve the scintillation properties established in \cite{Berns:2020ehg,3DET:2022dkw} and to reach the maximum possible material throughput velocity in order to obtain a high fill density of the PS volume by more forcefully spreading the melted polymer throughout the reflective cavity before it solidifies. This is achieved by elevating the heat block temperature (above the usual working temperature) such that the PS is melted to its core at the exit point. Simultaneously, the increased heat creep from the high-temperature heat block towards the whole extrusion system needs to be controlled, such that system clogging followed by the failure of the entire injection process is prevented. \\
The discretization of the model concluded in 321,545 nodes with a maximum polymer mesh size of 0.25~mm, a surface mesh size between one and two millimeters, three boundary layers between solid and fluid regions and a volume mesh filled by polyhedra (Fig.~\ref{fig:CFD model} b). 

Using the discretized model, the following components were determined: the filament-feeding-tube material and the heat break tube-thickness and material, both elements simulated for their heat shielding capabilities; the heat block material and temperature and the nozzle material, both responsible for the melting of the PS. The simulation was based on a laminar flow energy model using a pressure-based transient solver. The surrounding temperature (chamber temperature of the FDM printer) was set to 50°C, the circular cut-out in the heat block was set to a process temperature, simulating an FDM cartridge heater, and the polymer inlet velocity was varied to determine its maximum possible value. Table~\ref{tab:CFD mats} lists the investigated materials for each extrusion component to achieve a maximal material throughput.
The heat break is the essential element in the whole extrusion system because of its shielding function of the cold zone components from the heat creep by the heat block, thus guaranteeing a failure free process. The performance of the manufactured heat break based on the CFD analysis was measured with type ``k'' thermocouple from the manufacturer \textit{Fluke} and compared to the commercially available model.

\begin{table}[]
    \centering
    \begin{tabular}{c|c}
         \textbf{Extrusion component} & \textbf{Materials} \\
         \hline
         Polymer & Polystyrene \\
         Feeding tube & Stainless steel, PEEK \\
         Heat break & Stainless steel \\
         Heat break to heat block coupling & Stainless steel, brass, copper \\
         Heat block & Stainless steel, brass, copper \\
         Nozzle &  Stainless steel, brass, copper \\
    \end{tabular}
    \caption{Investigated material possibilities for each extrusion component to maximize polymer mass flow.}
    \label{tab:CFD mats}
\end{table}

%

%

\subsection{Detector setup and analysis method for cosmic data}
\label{sec:detector-cosmics}

The $5\times 5 \times 5$ \supercube~was instrumented with fifty readout channels, twenty-five on each view, as shown in Fig.~\ref{fig:SuperCube–cosmics-event-display}. 
Kuraray Y11 double-cladding 1~mm diameter WLS fibers \cite{kuraray-catalogue-wls} along the orthogonal X and Y directions allow for unique identification of the cubes crossed by a charged particle. 
WLS fibers absorb the blue light produced by the PS cube and isotropically emit photons in the green wavelength inside their core. The different refractive index of the outer cladding geometrically traps and guides the photons towards the SiPMs, thanks to an attenuation length of more than 4 meters. 
Hamamatsu Multi-Pixel Photon Counters (MPPCs) S13360-1325CS \cite{hamamatsu:mppc} were coupled with individual WLS fibers, on one of the two ends. 
The operating voltage was set to the one recommended by the supplier for each MPPC.
The coupling was ensured by black 3D printed optical connectors and enhanced by fixing the WLS fiber with EJ-500 optical glue \cite{EJ-500-glue} and pushing the MPPC toward the polished fiber end with a soft black foam acting like a spring.
The other end of the fiber was cut at $90^{\circ}$ and polished.
With a nominal photodetection efficiency of 25\%, the analogue charge signal of each MPPC is proportional, to the first order, to the number of photons produced in the PS cubes crossed by the corresponding WLS fiber.
The charge analogue signal of each MPPC is collected with micro-coaxial cables and read out by one CAEN FERS DT5202 front-end (FE) board \cite{caen-FERS-board}, used to digitize the signal. 
The charge value corresponding to the highest signal peak of each independent channel is measured and converted to analogue-to-digital converter (ADC) units. 
Then, from the measured number of ADC units, the number of photoelectrons (PE), i.e., the primary electrons originated via photoelectric effect by the visible photons impinging on the MPPC active area, is computed.
Such conversion was obtained with a large data sample collected by exposing the prototype to a $^{90}\text{Sr}$ radioactive source. The data could show a clear multi-peak structure, each one corresponding to a different number of PEs.
The MPPC gain could be extracted as the distance between adjacent peaks in units of ADC per PE
with a multi-Gaussian distribution fit 
and the position of the first three peaks could be obtained. 
Then, the peak positions were fitted with a linear function to extract the gain and pedestal. 
Since the recommended working bias voltage provided by the supplier was applied to each individual channel of the MPPC, the gain distribution was observed to be relatively uniform, averaging around 50 ADC/PE with a standard deviation of 1.21 ADC/PE. This procedure was iterated for each channel, employing its specific gain and pedestal value to convert the measured light yield into units of PE.
%
%
%
%
Examples of detected cosmic particles can be found in Fig.~\ref{fig:SuperCube–cosmics-event-display}. \\
The performance of the \supercube~produced with FIM was compared with a reference 2D matrix of cubes produced with cast polymerization. It consisted of a single layer of $5 \times 5$ optically-isolated PS $1~\text{cm}^3$ cubes, thus a geometry analogous to the \supercube~but two dimensional. 
In the same way as for the \supercube, both orthogonal views of the reference sample were read out with 
Kuraray Y11 WLS fibers, coupled with the same type of MPPC following the same procedure as described above. 
A more detailed description of the reference prototype as well as of its particle detection performance can be found in \cite{Boyarintsev:2021uyw}. 
The reference prototype was placed on top of the \supercube~and, read out with the same CAEN FERS DT5202 board, was used to trigger the cosmic particles and compare the light yield and crosstalk with the one of the \supercube.
%

Both the FIM and the cast polymerization prototypes were characterized by analyzing cosmic particle events and the respective light yield and crosstalk were compared. A total of about 
63
hours cosmic data was collected. 
Typical cosmic events are minimum ionising particles, crossing the prototype every few seconds with an expected energy loss in PS of about 1.8 MeV/cm. Given their angular distribution, approximately equal to $\cos^2 \theta_{\text{azim}}$ where $\theta_{\text{azim}}$ is the azimuth angle, the most common signature consists to the one of a vertical MIP leaving a track that starts from the top layer (the reference prototype) and ends at the bottom layer of the \supercube, as shown in Fig.~\ref{fig:SuperCube–cosmics-event-display}. 
The track 3D hits were reconstructed layer by layer, by determining the XY coordinate with the maximum light yield channel in both views. In each electronic channel a threshold of 500 ADC, i.e., about 10 PE, was applied. After the track reconstruction, a data set containing only vertical tracks was selected to have a particle path length of about 1 cm in each PS cube, hence a comparable energy loss. 
These tracks cross both the reference prototype and the \supercube~in the same column of cubes.
In Fig.~\ref{fig:SuperCube–cosmics-results}, a scintillation light yield of about 29 PE per cube is shown for both the \supercube~and the reference sample.
The cube-to-cube light crosstalk was derived from the light yield ratio between neighboring channels within the same layer perpendicular to the track direction. As shown in Fig.~\ref{fig:SuperCube–cosmics-results}, a slightly higher crosstalk was measured for the \supercube. However, it remained constrained at the few \% level, thus acceptable for particle tracking and calorimetry measurements.

\section{Data availability}


\section{Code availability}


\bibliographystyle{apsrev4-2}
\bibliography{Main}

\section{Acknowledgements}

Part of this work was supported by the SNSF grant PCEFP2\_203261, Switzerland. \\

\section{Author contributions}

D.S. and T.W. conceived the additive manufacturing method. 
T.W. implemented the method by designing and making the 3D printer components, tuning the 3D printer parameters and producing the final 3D printed sample. 
D.S. supervised the implementation process.
U.K. and B.L. built the experimental setup for cosmic particle data taking. 
B.L. ran the particle detection experiment and performed the data analysis.
C.J. and U.K. performed the measurements of reflector transmittance. 
U.K. and D.S. supervised the data taking and the data analysis.
Also A.Ru. supervised the data analysis.
A.B. and T.S. produced the scintillator filament used for the 3D printed prototype. 
S.B., E.B and S.H. provided some of the necessary equipment and discussed the results.
All the authors, part of the 3D printed DETector (3DET) R\&D collaboration, discussed and commented on the manuscript.\\




\section{Competing interests}

The authors declare no competing interests.

\end{document}